\documentclass[preprints,article,accept,moreauthors,pdftex]{Definitions/mdpi}

\usepackage{bm}
\usepackage{bbm}
\usepackage{listings}
\usepackage{subfig}
\usepackage{amsmath}
\usepackage{amssymb}
\usepackage{booktabs}
\usepackage{mathtools}
\usepackage{hyperref}
\usepackage{fourier}
\usepackage[T1]{fontenc}
\usepackage[]{avant}
\usepackage{longtable}

\firstpage{1} 
\pubvolume{}
\issuenum{1}
\articlenumber{1}
\pubyear{2024}
\copyrightyear{2024}
\history{}

\NewDocumentCommand{\expect}{ e{^} s o >{\SplitArgument{1}{|}}m }{%
  \operatorname{E}
  \IfValueT{#1}{{\!}^{#1}}
  \IfBooleanTF{#2}{
    \expectarg*{\expectvar#4}%
  }{
    \IfNoValueTF{#3}{
      \expectarg{\expectvar#4}%
    }{
      \expectarg[#3]{\expectvar#4}%
    }%
  }%
}
\NewDocumentCommand{\expectvar}{mm}{%
  #1\IfValueT{#2}{\nonscript\;\delimsize\vert\nonscript\;#2}%
}
\DeclarePairedDelimiterX{\expectarg}[1]{[}{]}{#1}

\newcommand*\diff{\mathop{}\!\kern0pt\mathrm{d}}

\newcommand{\Var}{\mathrm{Var}}

\newcommand{\VIX}{\mathrm{VIX}}
\newcommand{\VVIX}{\mathrm{VVIX}}
\newcommand{\Vega}{\mathrm{Vega}}

\Title{Heston vol-of-vol and the VVIX}
\Author{Jherek Healy}
\AuthorNames{Jherek Healy}

\address{jherekhealy@protonmail.com}
\abstract{The Heston stochastic volatility model is arguably, the most popular stochastic volatility model used to price and risk manage exotic derivatives. In spite of this, it is not 
necessarily easy to calibrate to the market and obtain stable exotic option prices with this model. This paper focuses on the vol-of-vol parameter and its relation with the volatility
of volatility index (VVIX) level. 
Four different approaches to estimate the VVIX in the Heston model are presented:
two based on the known transition density of the variance, one analytical approximation, and one 
based on the Heston PDE which computes the value directly out of the underlying SPX500. Finally we explore their use to improve calibration stability.}

\begin{document}
	\section{Introduction}
	
Under the Heston model \cite{heston1993closed}, the asset $X$ follows
\begin{subequations}
	\begin{align}
		\diff X(t) &= (r(t)-q(t)) X(t)dt+\sqrt{V(t)} X(t) \diff W_X(t)\,,\label{eqn:heston_X}\\ 
		\diff V(t) &= \kappa \left(\theta - V(t)\right) + \sigma \sqrt{V(t)} \diff W_V(t)\,,\label{eqn:heston_V}
	\end{align}
\end{subequations}
with $W_X$ and $W_V$ being two Brownian motions with correlation $\rho$, and $r, q$ the instantaneous growth and dividend rates.

The model, due to its affine properties, leads to an explicit characteristic function, enabling relatively fast pricing of European options. As a consequence, it has been extremely popular in the literature, as well as for real world use in banks,
in order to risk manage the price of exotic derivatives, such as Cliquet options \cite{gatheral2011volatility, guillaume2012calibration, feng2019cva}. 

While it is relatively straightforward to calibrate the model against a set of vanilla options,
it is difficult to obtain a stable, meaningful calibration. Indeed, a change of objective function used in the optimization, or the use of different tactics for calibration, 
such as reducing the number of parameters lead to vastly different prices for exotics. This is well described in \cite{guillaume2010use}, where the two of the model parameters
are set exogenously, $v(0)$ is set to the VIX value and $\theta$ is set from a moving window estimate of the historical VIX quotes. Some practitioners also fix the speed of mean reversion $\kappa$ exogenously. Yet another practice is to forget the consistency across 
deals, and  calibrate only against a single maturity $T$, setting $v_0 = \theta$ and a specific term-structure of $\kappa$ (\citet{clark2011foreign} recommends $5/T$).

The VVIX is an index measuring the volatility of the VIX index, which is itself a volatility index. The VVIX thus measures the vol of vol.
We thus may use the VVIX quote as vol-of-vol of the Heston model. But does the vol of vol $\sigma$ of the Heston model really corresponds to the VVIX? Does it help stabilizing the calibration in some sense? This paper intends to answer those questions, by analyzing the theoretical VVIX price in the Heston model.

\section{Approximate VVIX in the Heston model}
The VIX index corresponds to the square root of a variance swap replication for a maturity of 30 days, using a
simple discretization scheme, based on liquid vanilla option prices on the SPX500 index:
\begin{equation}
	\VIX(t,t+\Delta)^2  \Delta= -\left(1- \frac{F(t,t+\Delta)}{K^\star}\right)^2  + 2\sum_{i=1}^n w_i V_{\textmd{OTM}}(K_i, t, t+\Delta) \label{eqn:vix}\,,
\end{equation}
where $w_i = \frac{K_{i+1}-K_i}{2K_i^2}$ for $i=2,...,n-1$, $w_1 =  \frac{K_{2}-K_1}{2K_1^2}$,  $w_n =  \frac{K_{n}-K_{n-1}}{2K_n^2}$, $\Delta$ is 30 days (we will use the ACT/365 convention), $V_{\textmd{OTM}}(K_i, t, T)$ is the market price of an out-of-the-money option of strike $K_i$ and maturity $T$.
$K^{\star}$ is the first strike $K_i$  below  $F(t,t+\Delta)$.

We will approximate the VIX square by the true value of the expected 30-day variance in the Heston model
\begin{equation}
\VIX(t,T)^2 \approx   \frac{1}{T-t}\expect*{\int_t^T v(u) \diff u|t}\,.\label{eqn:vix_heston_expect}
\end{equation}
In the Heston model, the above expectation is known explicitly and reads
\begin{equation}
	\VIX_H(v(t),T-t)^2 =  \frac{1}{T-t}\expect*{\int_t^T v(u) \diff u|t} = \left(1 - \frac{1 - e^{-\kappa (T-t)}}{\kappa (T-t)}  \right) \theta + \frac{1-e^{-\kappa (T-t)}}{\kappa (T-t)} v(t)\,.\label{eqn:vix_heston}
\end{equation}

The buyer of a VIX future of maturity $T$ will receive at maturity, the VIX index value computed from option prices of maturity $T+\Delta$. Using the expected variance as VIX square (Equation \ref{eqn:vix_heston_expect}) leads to 
\begin{equation}
	F_{\VIX}(t,T) \approx  \expect*{ \sqrt{\VIX(T,T+\Delta)^2 }|t}\,.
	\end{equation}

Similarly, the VVIX corresponds to a variance swap replication for a maturity of 30 days, based on the VIX index. Options
on the VIX index of maturity $T$, are actually options on a  VIX future with maturity $T$, and thus measuring the volatility from $T$ to $T$ + 30 days, the 30 day forward volatility. 
It may be approximated by the log contract value \cite{bayer2016pricing}:
\begin{equation}
	\VVIX(t,T)^2 \approx  \frac{2}{T-t} \expect*{ \ln {F_{\VIX}(T,T)} - \ln F_{\VIX}(t,T)|t}\label{eqn:vvix_log}
	\end{equation}
	
For a CIR process, the transition law is known explicitly \cite{broadie2006exact} and reads 
\begin{equation}
v(t) = \frac{\sigma^2\left(1 - e^{-\kappa (t-u)}\right)}{4\kappa}\chi_d'^2\left(\frac{4\kappa e^{-\kappa (t-u)}}{\sigma^2 \left(1 - e^{-\kappa (t-u)}\right)}v(u) \right)\,,
\end{equation}
where $\chi_d'^2$ denotes the non-central chi-squared distribution with $d$ degrees of freedom where $d = \frac{4\kappa \theta}{\sigma^2}$. 

Let $C_1 = \frac{\sigma^2\left(1 - e^{-\kappa \Delta}\right)}{4\kappa}$, $\phi_{t,u}(z)$ be the probability density function of the non-central chi-squared distribution, combining Equation \ref{eqn:vix_heston} with Equation \ref{eqn:vvix_log}, we may thus compute the VVIX using a one-dimensional integration.
\begin{equation}
	\VVIX(t,T)^2 \approx   \frac{2}{T-t} \int_0^\infty   ( \ln \sqrt{ \VIX_H(v(T)= C_1 z,\Delta)^2 } - \ln F_{\VIX}(t,T)) \phi_{t,T}(z) \diff z \label{eqn:vvix_heston}
\end{equation}
with 
\begin{equation}
F_{\VIX}(t,T) = \int_0^\infty   \sqrt{ \VIX_H(v(T)=C_1 z, \Delta)^2 }  \phi_{t,T}(z) \diff z
\end{equation}
where the expectation inside each integral is given explicitly by Equation \ref{eqn:vix_heston}. The VIX future formula corresponds to the one given in \cite{zhang2006vix}.

When the Feller condition $\kappa \theta \geq \sigma^2$ does not hold, the integral is more challenging to calculate due to the explosion at zero. We found only minor discrepancies using Julia's QuadGK\footnote{\url{https://github.com/JuliaMath/QuadGK.jl}}.
Alternatively, \citet{zhu2012analytical} present a Fourier based approach to compute the expectation, using the characteristic function of the variance process, which may be more stable numerically and better performing.

In practice, the VVIX is calculated using only a discrete set of liquid options on the VIX:
\begin{equation}
	\VVIX(t,T)^2 (T-t)= -\left(1- \frac{F_{\VIX}(t,T)}{K^{V\star}}\right)^2 + \sum_{i=1}^n w_i V_{OTM}(K^V_{i}, t, T) \label{eqn:vvix_heston_exact}
\end{equation}
where $w_i = \frac{K^V_{i+1}-K^V_i}{2{K^V_i}^2}$ for $i=2,...,n-1$, $w_1 =  \frac{K^V_{2}-K^V_1}{2{K^V_1}^2}$,  $w_n =  \frac{K^V_{n}-K^V_{n-1}}{2{K^V_n}^2}$,
$K^{V\star}$ is the first strike $K^V_i$  below  $F_{VIX}$.

The price of VIX option can be computed using a one-dimensional integration, very much like for the continous VVIX approximation:
\begin{equation}
   V(K,t,T) =  \int_0^\infty   \max\left(\eta  \sqrt{ \VIX_H(v(T) = C_1 z, \Delta)^2 } - \eta K, 0\right) \phi_{t,T}(z) \diff z \label{eqn:vix_option}
\end{equation}
with $\eta=1$ for a call and $\eta=-1$ for a put option. A practical implementation will truncate the integration at the level $z$ where the intrinsic value is zero in order to keep the integrand smooth.

If we assume that the VIX option prices are given by Equation \ref{eqn:vix_option}, the continuous approximation (Equation \ref{eqn:vvix_heston}) is remarkably close to the more exact VVIX replication, as long as the lowest replication strike is small enough.

 A typical strike range is $[10, 65]$ for a VIX=14. In particular, the lowest strike is not that small
 compared to the VIX value. This creates a bias in the VVIX value, which may be particularly pronounced if the VVIX is large (Table \ref{tbl:vvix_truncation} with the first VIX option strike $K^V_1 = 5$ or $K^V_1 = 10$). 
 \begin{table}[H]
	\centering{
		\caption{Effect of the VIX option strike range truncation on the estimated VVIX value, compared to the log contract price, for different Heston parameters. \label{tbl:vvix_truncation}}
 \begin{tabular}{cccccccccccc}\toprule
Set & \multicolumn{5}{c}{Heston parameters} & $F_{\VIX}$ & Log  &\multicolumn{2}{c}{Replication} & Simple\\ 
&$v_0$ & $\kappa$ & $\theta$ & $\rho$ & $\sigma$ & & contract & $K^V_1 = 5$& $K^V_1 = 10$&\\
\midrule
I & 0.0236 & 0.2575 & 0.0849 & -0.7513 & 0.3150 & 15.4 & 105.4 & 105.2 (0.2\%) & 100.0 (5.1\%) &98 \\
II &0.0313 & 0.75 & 0.0678 & -0.7663& 0.7593 & 16.0 & 218.5 & 218.4 (0.0\%) & 193.0 (11.5\%) & 204\\
III & 0.0371& 3.4490& 0.0497&  -0.7558 & 1.7522 & 15.9 & 231.8 & 229.8 (0.9\%) & 225.9 (2.5\%) & 382 \\\midrule
IV & 0.0538 & 0.6431 & 0.0880 & -0.7010 & 0.6159 & 22.2 & 143.7  & 143.8 &139.8 &127\\ 
V  &0.0440 & 0.75 & 0.0998 & -0.7410 &0.7654 & 19.9 & 184.8 & 184.8 & 176.8 &171\\
VI & 0.0397 & 4.6705 & 0.0696 & -0.7149 & 2.0640 & 18.9 & 196.7 & 196.5  & 196.7& 376\\ 
\bottomrule
 \end{tabular}}
 \end{table}
 We also derive a simple approximation in Appendix \ref{sec:approx}, which does not require any numerical integral calculation, but is not particularly accurate, especially for large speeds of mean reversion $\kappa$ (column named "Simple" of Table \ref{tbl:vvix_truncation}).

\section{Exact VVIX with the Heston PDE}
The VIX itself is also a discrete replication of out-of-the-money 30-days vanilla options on SPX500, which is not taken into account in the single replication approach presented in the previous section.
In general, there is much less issue with the lowest strike, and SPX500 options are liquid across a wider range of strikes.
A PDE approach allow us to compute the VVIX based SPX500 options only,
by combining the discrete replication of the VVIX and of the VIX. The full Heston model is used, not only the variance process, there is no approximation involved. Furthermore, the technique is applicable to any stochastic volatility model.

The Heston PDE for the price of a financial derivative $f$ reads
	\begin{equation}
		\frac{\partial f}{\partial t} = \frac{v x^2}{2}\frac{\partial^2 f}{\partial x^2} + \rho\sigma x v \frac{\partial^2 f}{\partial x \partial v} + \frac{\sigma^2 v}{2} \frac{\partial^2 f}{\partial v^2} + (r-q)x\frac{\partial f}{\partial x} + \kappa(\theta - v)\frac{\partial f}{\partial v} - r_C f\,,\label{eqn:heston_pde}
	\end{equation}
	for $0 \leq t \leq T+\Delta$, $x > 0$, $v > 0$.	 We allowed for a distinct discounting rate $r_C$, which will be set to zero to obtain undiscounted prices of $f$.
	We use the  boundary conditions and central finite difference discretization described in \cite{lefloc2023instabilities, lefloch2021pricing}. 

We consider $f$ to be of dimension $2n$ where $n$ is the number of strikes used in the replication. The first $n$ are call options and the next $n$ are put options.

\subsection{Initial condition}
The initial condition at $t=T+\Delta$ reads 
\begin{align}
f_i(x,v,T+\Delta) = \max(x-K_i,0)\,, \quad f_{n+i}(x,v,T+\Delta) = \max(K_i-x,0)\,,
\end{align}
for $x \geq 0$, $v \geq 0$, $i=1,...,n$. 

\subsection{Continuity condition}
At $t=T$, we calculate the VIX options in three steps:
\begin{enumerate}[label=(\roman*)]
	\item We compute the SPX500 forward estimate at each point. To do so, we perform a linear regression on the call-put parity relation, that is we solve
	$\bm{H}^T \bm{H}  \bm{\beta} = \bm{H}^T \bm{G}$
	where $\bm{H}$ is a (n,2)-matrix with $H_{i,1}=K_i$, $H_{i,2}=1$ and $\bm{G}$ is a $n$-dimensional vector with ${G}_{i} = f_i(x,v,T)-f_{n+i}(x,v,T)$. Then the forward at each point is $F(x,v) = \beta_2(x,v)$.
	\item We compute the VIX price at each point. We apply Equation \ref{eqn:vix}, using $F$, $f_i$ and $f_{n+i}$. This leads to $\VIX(x,v)$.
	\item We update $f$ to be the value of VIX options for each strike:
	\begin{align}
		f_i(x,v,T) = \max\left(\VIX(x,v)-K^V_i,0\right)\,, \quad f_{n+i}(x,v,T) = \max\left(K^V_i-\VIX(x,v),0\right)\,,
		\end{align}for $x \geq 0$, $v \geq 0$, $i=1,...,n$. 
\end{enumerate}

At $t=0$ we end up with the values of VIX options replicating the VVIX, we then apply (i) and (ii) to those in order to obtain the VVIX at each point. We then use a bicubic spline to find the VVIX value at the current SPX500 spot price and the Heston initial variance $v(0)$.

\section{Numerical Results}
\subsection{Calibration}
The calibration of the Heston model consists in finding the parameters which  minimize the difference between the model option prices $\hat{\bm{V}}$ and the market option prices $\bm{V}$. There are many subtleties in practice:
\begin{itemize}
	\item Which market option prices do we include? Do we include all, or only liquid prices?
	\item  The model is also often calibrated against a prior representation of market option prices, typically an implied volatility matrix corresponding to a fixed range of maturities and strikes. This prior representation will typically not have any liquidity information.
	\item Should we select call option prices, put option prices or only out-of-the-money option prices? 
	\item Which measure of difference do we choose? $\ell^{\infty}$, the maximum absolute error or $\ell^{2}$, the sum of square differences? Should it be weighted in some ways? Should it be applied to implied volatilities or to raw prices?
\end{itemize}

We will adopt the $\ell^2$ norm since it enables the use of fast local minimizers such as the Gauss-Newton of \citet{klare2013gn}. As initial guess, we use the result of a global minimizer \cite{storn1997differential}.

Let $T_i$, $K_i$ be the maturities and strikes for each of the $n$ market options considered, the various choices may be reduced to minimize a sum of weighted square differences in out-of-the-money option prices:
\begin{equation}
	\bm{\xi}_{\textmd{calibrated}} = \min_{\bm{\xi}}\left[\sum_{i=1}^n w_i^2 \left(\hat{V}(K_i,T_i, \bm{\xi}) - V(K_i,T_i)\right)^2\right]\,,
\end{equation}
where $\bm{\xi}$ represent the vector of parameters to calibrate. For the full Heston model, $\bm{\xi}= (v_0, \kappa, \theta, \rho, \sigma)$.

If we let $\bm{w}=1$, which is often used in the literature, the very out-of-the-money option prices, where the liquidity is not always good, will effectively be ignored as their price will be negligible compared to the closer to the money option prices. Furthermore, it will put more emphasis on longer maturities than on shorter ones. 

We may also let $w_i = \frac{1}{B(0,T_i)}$ where $B(0,T_i)$ is the discount factor to maturity $T_i$, which is equivalent to working with undiscounted prices and $\bm{w}=1$.

If we let $w_i = \frac{1}{\min\left( \Vega_{F}, \Vega(K_i,T_i)\right)}$, where 
\begin{equation*}\Vega = B(0,T_i)F(0,T_i)\phi\left(\frac{1}{\sigma_i\sqrt{T_i}}\ln \frac{F(0,T_i)}{K_i} + \frac{1}{2}\sigma_i\sqrt{T_i}\right)\sqrt{T_i}\end{equation*} is the market Black-Scholes Vega, and $F(0,T_i) = X(0) e^{\int_0^{T_i} r(s)-q(s)\diff s}$ is the forward to maturity $T_i$, we end up with a normalized difference which is close to the error in implied volatilities (see \cite{lefloch2019model1} for a justification). In addition, the floor $\Vega_{F}$ allows to relax the fitting of illiquid out-of-the-money option whose Vegas are typically very small. We will use $\Vega_{F} = 10^{-2}$ in our numerical tests.

Table \ref{tbl:vvix_truncation} presents the resulting parameters for different choices of $\bm{w}$ when we calibrate towards SPX500 options as of October 8, 2024: Set I is equal weights, Set III is inverse vega, and Set II is inverse vega with a exogenous $\kappa=0.75$. Set IV, V, VI correspond to the calibrations as of March 15, 2021.
We notice that depending on the choice of objective function, the calibrated parameters lead to vastly different  levels of VVIX, some of which are not compatible with the current market conditions. The VVIX trades mostly between 65 and 200, and typically hovered around 100 in 2023 and 2024.  On October 8, 2024, it was 114.82, and on March 15, 2021, it was 110.49.

This suggests that the log contract formula (or even the simple approximation) may be good enough to calibrate Heston. We will however firstly estimate the impact of the true double replication on the VVIX estimate, in comparison to the single replication.


\subsection{PDE} 
We use the Runge-Kutta-Gegenbauer (RKG) second-order explicit finite-difference method \cite{skaras2021super,lefloc2023instabilities} to price the double replication of the VVIX on the Heston PDE. As of December 2024, the SPX is close to 6000 and the strike range for liquid SPX options of maturity 2 months lies between 2500 and 8000, which corresponds to a range from 40\% to 140\% of the spot price. We will use this replication range in our numerical tests.

We consider the two sets of parameters with largest vol-of-vol $\sigma$, and price the VVIX, successively doubling the grid density to show the convergence (Table \ref{tbl:pde_convergence}). We increased the boundary in the asset dimension $X$ to six standard deviations in order to avoid perturbations related to the boundary.
\begin{table}[H]
	\caption{VVIX price by double replication in the Heston PDE, discretized with the RKG scheme, doubling the grid size. $N$ is the number of time-steps, $M$ the number of steps in the $X$ axis, $L$ the number of steps in the $v$ axis.\label{tbl:pde_convergence}}
	\centering{
\begin{tabular}{rrrccc}\toprule
N & M & L & \multicolumn{2}{c}{Set II} &\multicolumn{1}{c}{Set III}  \\
& & & $K_1^V = 5$  & $K_1^V = 10$ & $K_1^V = 10$\\  \midrule
25 & 12 & 16 &197.96& 185.08 &217.04\\
50 & 25 & 30 &211.54& 189.63 &225.26\\
100 &50 & 60 &214.11& 191.07 &223.89\\
200 &100&120 &217.53& 192.62 &224.23\\
400 &200&240 &218.00& 192.78 &224.66 \\\bottomrule
\end{tabular}}
\end{table}
The VVIX double replication also enables us to evaluate the effects of truncating the range of SPX option strikes used in the VIX replication. As expected, the difference between the PDE double-replication values and the single replication (Table \ref{tbl:vvix_truncation}) is small, because the SPX500 options are liquid on a wide enough strike range.



\subsection{VVIX level versus vol-of-vol}
\begin{figure}[h!]
	\centering{
		\subfloat[][Set I.]{\includegraphics[width=0.48\textwidth]{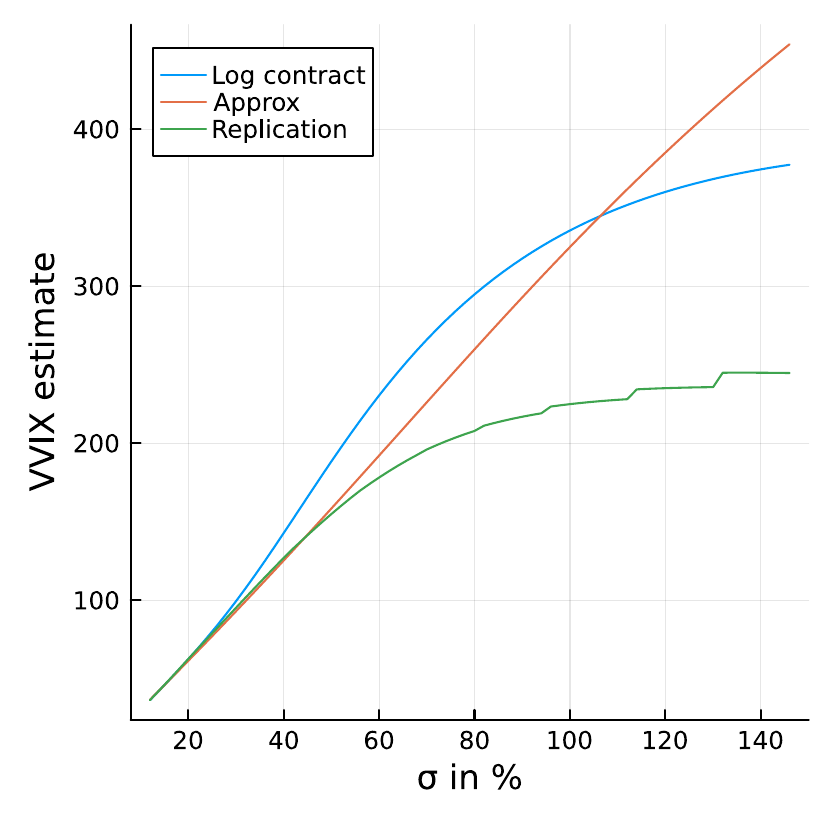}}
		\subfloat[][Set III.]{\includegraphics[width=0.48\textwidth]{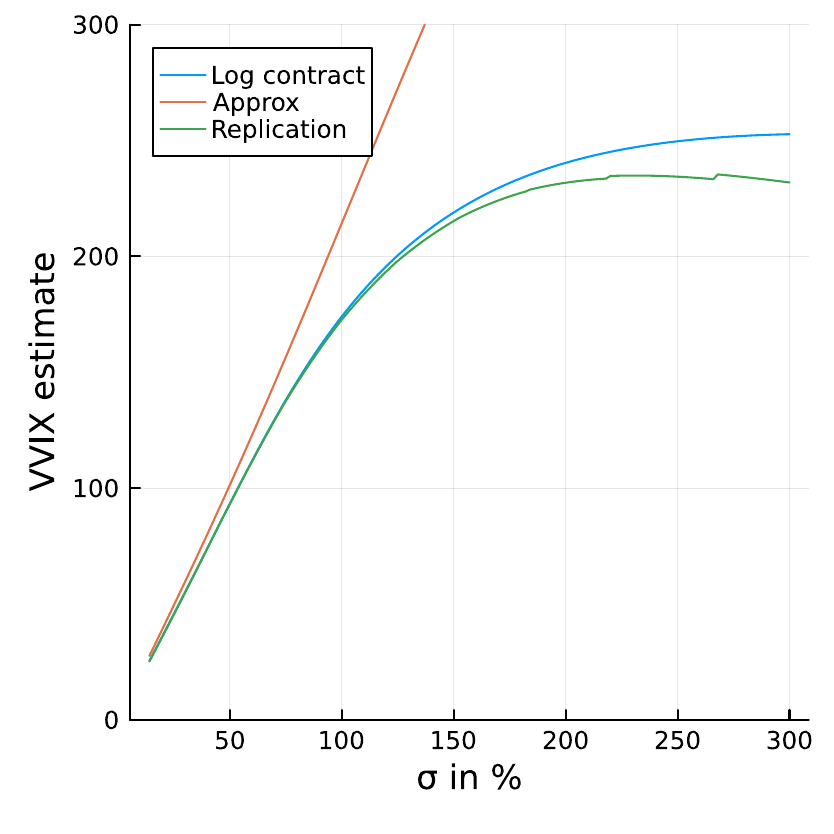}}
	}
	\caption{VVIX varying the vol-of-vol but keeping the other parameters constant.}
	\label{fig:vvix_set_iv}
\end{figure}

We have seen in Tables \ref{tbl:vvix_truncation}, \ref{tbl:pde_convergence} that the vol-of-vol is different from the VVIX level, a vol of vol of 30\% may translate into a VVIX of 100\%.
In Figure \ref{fig:vvix_set_iv}, we take a closer look at how the VVIX evolves when we vary the vol-of-vol but keep the other Heston parameters constant, based on the parameters for Set I and Set III.
The approximation is inaccurate as soon as the vol-of-vol is larger than 50\%. In Set I, which has a small $\kappa$, the difference between the log-contract and the replication is large when the VVIX estimate is larger than 200. This is not the case for Set III, where the $\kappa$ is ten times larger. For Set III, the VVIX seems to reach a maximum of around 240. In particular, a vol-of-vol over 125\% leads to a VVIX between 200 and 230. On Set I, the same VVIX range is covered by a vol-of-vol  over 70\%. Thus, the vol-of-vol is not the single factor that will have a impact on the VVIX, the $\kappa$ must be considered as well.

\begin{figure}[h!]
	\centering{
		\subfloat[][Equals weights (Set I).]{\includegraphics[width=0.48\textwidth]{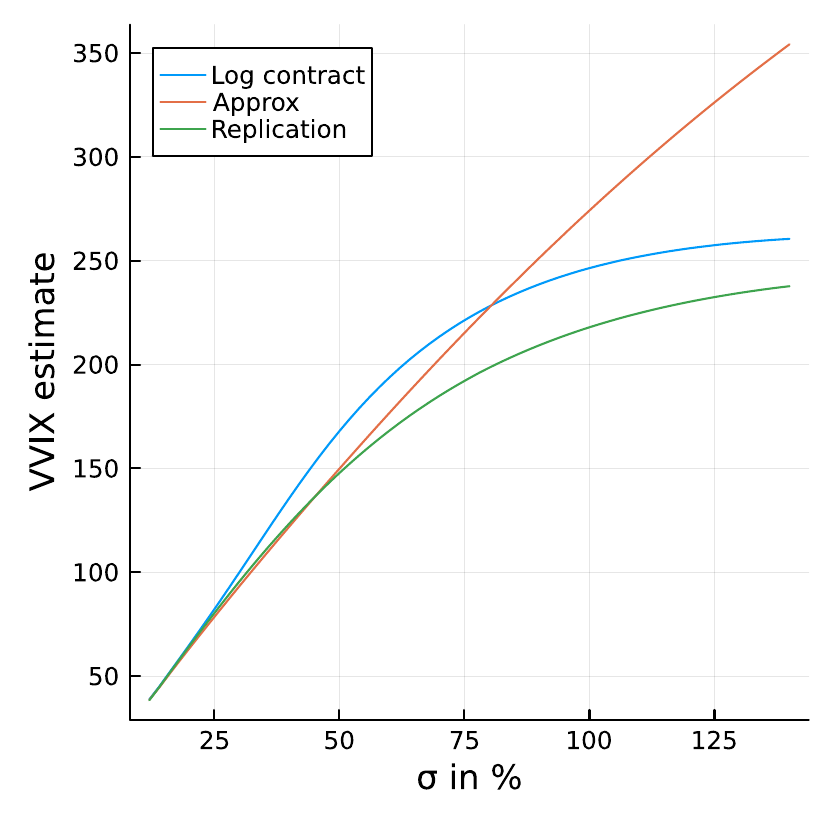}}
		\subfloat[][Inverse Vega weights (Set III).]{\includegraphics[width=0.48\textwidth]{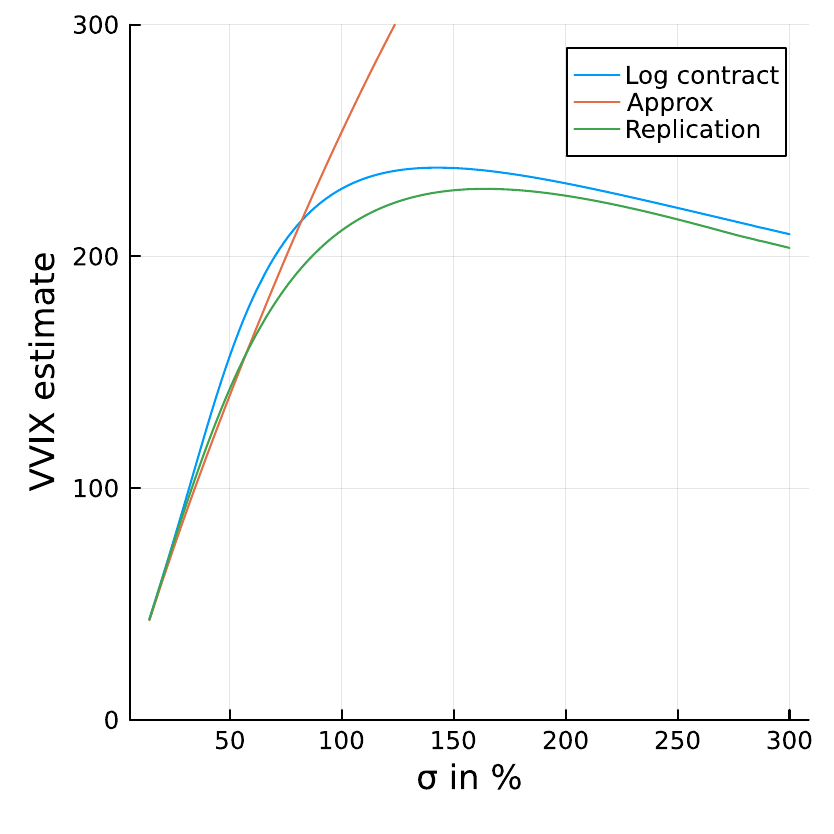}}
	}
	\caption{VVIX varying the vol-of-vol and recalibrating the other parameters against the market.}
	\label{fig:vvix_set_iv_recalibrated}
\end{figure} 
It is interesting to look at the same example, but with a recalibration of the other parameters (beside the $\sigma$) to the market. We calibrate to all expiries, even though the VVIX is actually really sensitive to maturities up to two months.
Figure \ref{fig:vvix_set_iv_recalibrated} shows that the VVIX of 200 is reached at around $\sigma=75\%$ for the two objective functions considered (corresponding to Sets I and IV). The recalibration appears to make the choice of $\sigma$ much more impactful.


\subsection{Calibration using VVIX}
Two different strategies are possible include the VVIX in the Heston calibration:
\begin{enumerate}[label=(\alph*)]
	\item An additional term in the least squares fit corresponding to a weighted difference between the model VVIX and the market VVIX.
	\item Or remove the vol-of-vol $\sigma$ from the set $\bm{\xi}$  and solve for it from the market value and the other Heston parameters. 
\end{enumerate}

Strategy (a) is more flexible, but requires to find a practical value for the VVIX associated weight. Strategy (b) does not have a weight to define and may decrease the calibration time. It will however require to be able to solve, meaning to have a monotonic function of $\sigma$. This excludes, a priori the replication approach (see Figure \ref{fig:vvix_set_iv}).

It turns out that strategy (a) fails in many situations, even with the log contract VVIX valuation, also because of  the non-monotonic relation with $\sigma$ (Figure \ref{fig:vvix_set_iv_recalibrated}): it may push $\sigma$ very high and $\kappa$ low. Even if the $\kappa$ is fixed exogenously, then the $\theta$ may be pushed unrealistically high. Both $\theta$ and $\kappa$ would be needed to be set exogenously to stabilize the calibration. As illustrating example, we calibrate Heston with inverse Vega weights and set the VVIX level to 200.0 with a weights of 0.1, using the log contract approach to match the VVIX, as of October 8, 2024. We obtain $v_0 = 0.04, \kappa = 7.17, \theta=0.05, \rho=-0.76, \sigma=3.09$. Both $\sigma$ and $\kappa$ are significantly larger than without the VVIX observation. The satisfying solution would be $v_0 = 0.03, \kappa=0.95, \theta=0.06, \rho=-0.79, \sigma=0.71$. With the replication approach strategy (a) fails even more often and more significantly. It however works satisfyingly with the simple approximation. The main cause is the non monotonicity of the VVIX theoretical price as a function of the vol-of-vol in the Heston model. One would need some additional constraint to put emphasis on the solution with lower vol-of-vol compared to the one with higher vol of vol. In a sense, this is precisely what the simple approximation accomplishes.

In strategy (b), a simple Newton solver to find $\sigma$ for a given VVIX level worked well enough and the calibration was 50\% faster than with strategy (a).

\begin{table}[H]
	\centering{
		\caption{Heston calibration on vanilla option prices and VVIX using strategy (b). Sets I and III correspond to uniform and inverse Vega weights calibration as of October 8, 2024. Sets IV and VI are as of March 15, 2021.\label{tbl:vvix_calibration}}
 \begin{tabular}{ccccccccccc}\toprule
Set & Simple & Log  &Replication& \multicolumn{5}{c}{Heston parameters} \\ 
& approx. & contract & $K^V_1 = 10$& $v_0$ & $\kappa$ & $\theta$ & $\rho$ & $\sigma$ \\
\midrule
I & 114.82 & 126.7 & 115.7 & 0.024 & 0.317 & 0.081 &-0.724 &0.374\\
III & 114.82 & 127.1 & 118.5 & 0.028 & 0.426 &0.069 &-0.842& 0.406\\\midrule 
I & 200.00 & 210.7 & 179.8 & 0.026 & 0.650 & 0.072& -0.672 &0.683\\
III & 200.00 & 207.1 & 185.5 & 0.032 & 1.045 & 0.058 & -0.788 &0.759\\\midrule 
IV & 110.49 & 122.6 & 119.8 &  0.052 &0.502&0.092&-0.703&
0.522\\
VI & 110.49 & 118.8 & 116.3 &  0.042 & 0.658 &0.086& -0.758&0.486\\\midrule 
IV & 200.00 & 210.9 & 204.1 &  0.062 & 1.328 &0.080 &-0.697 &1.049\\
VI & 200.00 & 188.3 &  183.8 & 0.043 & 1.688 &0.076 &-0.736&0.939\\ 
\bottomrule
 \end{tabular}}
 \end{table}
 Table \ref{tbl:vvix_calibration} shows a reduction of the discrepancy between the different choices of weights, significantly so when the VVIX level is below 150, and less so for a VVIX around 200.

\section{Conclusion}
We have explored the use of the VVIX index for calibration in an attempt to solve the major issue of inconsistent calibration and make the Heston model more practical.

The VVIX index value can not be used directly as vol-of-vol of the Heston model. The vol-of-vol is typically smaller
than the VVIX for moderate values of the VVIX. The truncation(s) involved in the VVIX replication(s) does not have a  large impact on the real VVIX value, with the exception of the combination of a low VIX (below 15) and large VVIX (above 150) where the difference can reach a 10\%.

Using the replication price or even the log contract price as part of the calibration proved to be challenging, because the relationship between the vol-of-vol and the theoretical VVIX value is not monotonic. A simple analytical approximation was however found to work well, and allowed to reduce the discrepancy due to the various choices in the calibration significantly. This approximation is not so accurate for large vol-of-vol values. The search for better approximation of the VVIX under Heston may be worthwhile. 

We may also wonder if the selection towards lower speeds of mean reversion and lower vol-of-vol due to the approximation is really justified in the context of a large market VVIX (such as 200\%).

Finally, we would expect a more direct relationship for models with a lognormal stochastic volatility, such as the SABR or Bergomi models.

\externalbibliography{yes}
\bibliography{heston_vvix.bib}
\appendixtitles{no}

\appendix
\section{Simple approximation for the VVIX in Heston}\label{sec:approx}
We have
\begin{equation*}
	\Var\left[\sqrt{\VIX_H(v(T),\Delta)^2}\right] =  \expect*{\VIX_H(v(T),\Delta)^2 |t} - \expect*{\sqrt{\VIX_H(v(T),\Delta)^2}}^2
\end{equation*}

Then we use a Taylor expansion of  $\sqrt{\VIX_H(v(T),\Delta)^2}$ around $\expect{\VIX_H(v(T),\Delta)^2|t}$ to obtain:
\begin{align*}
	\expect*{\sqrt{\VIX_H(v(T),\Delta)^2}} &\approx \sqrt{\expect*{\VIX_H(v(T),\Delta)^2|t}} + \frac{ \expect*{\VIX_H(v(T),\Delta)^2 - \expect*{\VIX_H(v(T),\Delta)^2|t}|t} }{2\sqrt{\expect*{\VIX_H(v(T),\Delta)^2|t}}}\\&-\frac{  \expect*{\left(\VIX_H(v(T),\Delta)^2 - \expect*{\VIX_H(v(T),\Delta)^2|t}\right)^2|t} }{8 \left(\expect*{\VIX_H(v(T),\Delta)^2|t}\right)^{\frac{3}{2}}} 
\end{align*}
This leads to 
\begin{align}
	\Var\left[\sqrt{\VIX_H(v(T),\Delta)^2}\right] \approx  \frac{ \expect*{\left(\VIX_H(v(T),\Delta)^2 - \expect*{\VIX_H(v(T),\Delta)^2|t}\right)^2|t}  }{4 \left(\expect*{\VIX_H(v(T),\Delta)^2|t}\right)} 
\end{align}

On one hand, we have \cite{gatheral2011volatility}
\begin{equation}
\expect{v(T)|t} = (v(t)-\theta)e^{-\kappa (T-t)} + \theta \label{eqn:eVT}
\end{equation}
And we thus know $\expect*{\VIX_H(v(T),\Delta)^2|t}$ explicitly using Equation \ref{eqn:eVT} in Equation \ref{eqn:vix_heston}. 

On the other hand, the dynamic of $U=v^2$ under Heston reads
\begin{equation}
\diff U = 2\kappa(\theta+ \sigma^2 - \sqrt{U})\sqrt{U} \diff t + 2 \sigma^2 U^{3/4} \diff W_v\,,
\end{equation}
which leads to 
\begin{equation}
\expect{v^2(T)|t} = \left[\frac{\sigma^2}{2\kappa} + \theta + e^{-\kappa (T-t)}\left(\frac{\sigma^2}{2\kappa} + \theta-v_0 \right)\right]^2 \label{eqn:eV2T}
\end{equation}
as the expectation removes the stochastic term $\sigma^2 U^{3/4} \diff W_v$ since it is a martingale with zero expectation.

We use Equations \ref{eqn:eVT} and \ref{eqn:eV2T} in the square of Equation \ref{eqn:vix_heston}, to obtain $\expect*{\VIX_H(v(T),\Delta)^4|t}$.

Then a simple lognormal approximation for the VVIX reads
\begin{equation}
\VVIX_{A} = \sqrt{\ln\left( 1 + \frac{ \Var\left[\VIX_H(v(T),\Delta)\right] }{ \expect{\VIX_H(v(T),\Delta)^2|t }}\right)} / \sqrt{\Delta}\,. \label{eqn:vvix_simple}
\end{equation}

	

\end{document}